\documentclass[%
reprint,
superscriptaddress,
groupedaddress,
amsmath,amssymb,
aps,
prb,
]{revtex4-2}

\usepackage{xcolor}
\usepackage{multirow}
\usepackage{graphicx}
\usepackage{dcolumn}
\usepackage{bm}
\usepackage{hyperref}
\usepackage{float}

\usepackage{booktabs}
\usepackage{amsmath}

\setcitestyle{super}

\begin{document}

\title{First-Principles Calculation of Hubbard U for Terbium Metal under High Pressure}

\author{Logan A. Burnett}
\email{loganb14@uab.edu}
\affiliation{Department of Physics, University of Alabama at Birmingham, Birmingham, Alabama 35294, USA}

\author{Matthew P. Clay}
\affiliation{Department of Physics, University of Alabama at Birmingham, Birmingham, Alabama 35294, USA}

\author{Yogesh K. Vohra}
\affiliation{Department of Physics, University of Alabama at Birmingham, Birmingham, Alabama 35294, USA}

\author{Cheng-Chien Chen}
\email{chencc@uab.edu}
\affiliation{Department of Physics, University of Alabama at Birmingham, Birmingham, Alabama 35294, USA}

\date{\today}%

\begin{abstract}
Using density functional theory (DFT) and linear response approaches, we compute the on-site Hubbard interaction $U$ of elemental Terbium (Tb) metal in the pressure range $\sim 0-65$ GPa. The resulting first-principles $U$ values with experimental crystal structures enable us to examine the magnetic properties of Tb using a self-consistent DFT+U method. The lowest-energy magnetic states in our calculations for different high-pressure Tb phases -- including hcp, $\alpha$-Sm, and dhcp -- are found to be compatible with the corresponding magnetic ordering vectors reported in experiments. The result shows that the inclusion of Hubbard $U$ substantially improves the accuracy and efficiency in modeling correlated rare-earth materials. Our study also provides the necessary $U$ information for other quantum many-body techniques to study Tb under extreme pressure conditions.
\end{abstract}


\maketitle

\section{Introduction}
Rare-earth elements or the lanthanides form a chemically similar series due to the progressive filling of $4f$ electrons across the series. This similarity extends to the atomic structures, as the pure lanthanide metals adopt a common sequence of close-packed structures with increasing pressure or decreasing atomic number~\cite{samudrala_2013}. With a few exceptions, the lanthanides are paramagnetic at room temperature and ambient pressure, and magnetically ordered at low temperatures~\cite{gimaev_2021,koehler_1965}. The resulting magnetic structures can be complex, showing multiple phase transitions with decreasing temperature and/or increasing pressure. Due to challenging experimental conditions, little is known about the magnetic structures of the lanthanides at high pressures. Advances in theory and modeling can offer a promising path to fill this knowledge gap.

In this work, we focus on the heavy lanthanide terbium (Tb), which crystallizes in the hexagonal close-packed (hcp) structure at ambient conditions and undergoes the common lanthanide structural sequence with increasing pressure. Tb transforms into a samarium-like structure ($\alpha$-Sm) at roughly 4 GPa of applied pressure, a double hexagonal close-packed (dhcp) structure at 16 GPa, a distorted face-centered cubic (hR24) structure at 30$\pm$3 GPa, and an orthorhombic structure (oF16) at 51$\pm$2 GPa \cite{samudrala_2013,mcmahon2019structure}. At ambient pressure, Tb undergoes two magnetic phase transitions with reducing temperature, first to a helical antiferromagnetic (AFM) structure at 229 K and then to a ferromagnetic (FM) structure at 221 K, both with the magnetic moments residing in the basal plane \cite{koehler_1965}. The neutron diffraction study by Kozlenko et al. \cite{kozlenko2021pressure} showed that in the $\alpha$-Sm phase, the magnetic ordering temperatures are reduced, ferromagnetism is suppressed, and two AFM phase transitions occur when the temperature decreases. Additional insights about magnetism in the dhcp and hR24 phases were found in the combined neutron and x-ray diffraction study by Clay et al. \cite{clay_2023}. Their neutron diffraction spectrum at 20 GPa in the dhcp phase indicated a low-temperature AFM structure, but there were not enough data to model the structure.
To account for the complex series of phases and their transitions, electronic correlation effect is crucially needed. 

The properties of rare-earth materials are dominated by $4f$ electrons, especially when their energies are close to the Fermi level. The $4f$ orbitals are spatially more localized, resulting in a less dispersive band and stronger electron interaction. An explicit inclusion of the Hubbard $U$ parameter is typically considered in the context of the Hubbard model~\cite{hubbard1963} or in an Anderson-Kondo model~\cite{anderson1961localized, kondo1964resistance}. In the former, the $4f$ electrons hop directly between different rare-earth lattice sites; in the latter, $4f$ orbitals would hybridize with itinerant $5d/6s$ conduction bands~\cite{chen2019probing}. In either case, an on-site repulsive interaction between doubly occupied $4f$ orbitals is included, with the $U$ parameter controlling the interaction strength.

In the literature, a wide range of $U \sim 3 - 9$ eV have been considered for Tb compounds~\cite{lebegue2006multiplet,peters2014treatment,peters2016magnetism, locht2016standard,mcmahon2019structure}. Since the electronic and magnetic properties in rare-earth materials can depend subtly on $U$~\cite{PhysRevB.108.075150,shi2023absence}, establishing the proper $U$ values for Tb metal under pressure would be an important first task. The $U$ value may be obtained by fitting the experimental magnetic moment sizes and/or electronic band structures to first-principles calculations utilizing the density functional theory with Hubbard correction (DFT+U)~\cite{PhysRevB.44.943, PhysRevB.52.R5467, VladimirIAnisimov_1997}, which is a relatively simple and efficient numerical approach to account for electron correlation effects. However, details of the magnetic phases or $4f$-orbital energies are often not readily available in high-pressure experiments. On the other hand, it is possible to determine $U$ solely from {\it ab initio} calculations, for example using constrained random-phase approximation~\cite{PhysRevB.70.195104, PhysRevB.74.125106}. This method can be accurate but computationally involved, with the additional complication and challenge of wannierizing entangled electronic bands~\cite{PhysRevB.65.035109,PhysRevB.80.155134,PhysRevB.83.121101,PhysRevB.86.165105}.

In this paper, we determine the effective first-principles $U$ values of Tb in the pressure range $\sim 0-65$ GPa, using the linear response ansatz by Cococcioni and de Gironcoli~\cite{cococcioni2005linear}. We further utilize the resulting $U$ values with the experimental crystal structures to study different magnetic states. The lowest-energy states in our self-consistent DFT+U calculations have magnetic ordering vectors consistent with the low-temperature high-pressure experiments, showing the promise of the employed linear response and DFT+U approaches.
The $U$ values we obtained also can provide the input and benchmark for other advanced quantum many-body techniques, based on dynamical mean-field theory~\cite{lebegue2006multiplet, peters2014treatment, locht2016standard, mcmahon2019structure} or other numerical approximations~\cite{hughes2007lanthanide, kozlenko2021pressure}. Our paper is thereby important for modeling rare-earth materials, especially Tb, in extreme environments.

The rest of the paper is organized as follows: Sec. \ref{methods} discusses the details of first-principles density functional theory (DFT) calculations. Sec. \ref{results} presents the resulting Hubbard $U$ values versus external pressure, the DFT+U energies for different high-pressure phases, as well as the magnetic moments and charge occupation of corresponding DFT+U ground states. Sec. \ref{conclusion} summarizes the paper with an outlook on future work.

\section{Methods}\label{methods}
Our DFT calculations are based on the plane-wave pseudopotential code Vienna Ab initio Simulation Package (VASP)~\cite{PhysRevB.54.11169, kresse1996efficiency}.
We opt for the Perdew-Burke-Ernzerhof generalized gradient approximation (PBE-GGA) functional~\cite{PhysRevLett.78.1396} and the projector augmented wave basis~\cite{PhysRevB.50.17953, kresse1999ultrasoft}. PBE-GGA is part of the wider family of GGA functionals, which generally offer improved accuracy over the local density approximation for the electron exchange-correlation energy, making it a logical choice for this study~\cite{ZIESCHE1998122}.
To prevent introducing extra errors in the crystal structure information for VASP input, we directly utilize crystallographic data from high-pressure experiments: The hcp and $\alpha$-Sm structures are from Kozlenko et al.~\cite{kozlenko2021pressure}; the dhcp and hR24 structures are from Clay et al.~\cite{clay_2023}; for oF16, the structure is from McMahon et al.~\cite{mcmahon2019structure}.

In general, converging DFT calculations for correlated materials requires careful consideration of various parameters within the VASP framework, due to the complexities associated with localized $d$ or $f$ orbitals~\cite{Emery2017}. In our calculations, the linear mixing parameters for the electronic density updates between DFT iterations play crucial roles. In particular, we have used the linear mixing parameter AMIX = 0.2, the cutoff wave vector for the Kerker mixing scheme~\cite{PhysRevB.23.3082} BMIX = 0.0001, the linear mixing parameter for the magnetization density AMIX\textunderscore MAG = 0.8, and the cutoff wave vector for the Kerker mixing scheme~\cite{PhysRevB.23.3082} for the magnetization density BMIX\textunderscore MAG = 0.001. These mixing parameters can be optimized to significantly improve the stability and efficiency of the self-consistent field iterations~\cite{PhysRevB.23.3082}. The optimal AMIX depends very much on the system, and here it is set to half of the default value. The BMIX parameter is set close to zero to speed up convergence, but without comprising stability. 
Moreover, we find that a relatively high plane-wave cutoff energy ENCUT is necessary to accurately model the heavy element Tb under pressure. We used ENCUT = 650 eV to help stabilize the calculations and reduce the required number of steps for reaching convergence. 

We have performed convergence tests with respect to ENCUT, k-point discretization, and break conditions. The calculations of the Hubbard $U$ achieved convergence with ENCUT = 650 eV and a stringent global break condition of $10^{-8}$ eV in the electronic self-consistent loop. A fine $7\times 7\times 7$ k-grid is adopted for the largest 54-atom Tb supercell, yielding a high KPPRA (k-points per reciprocal atom) of 18,522. The DFT ground-state calculations for various crystal structures provide the necessary input for subsequent linear-response calculations~\cite{cococcioni2005linear} to determine the $U$ values in the pressure range $\sim 0-65$ GPa.

After computing the Hubbard $U$ values, we further perform self-consistent DFT+U calculations to study the electronic and magnetic properties of Tb for the hcp, $\alpha$-Sm, and dhcp phases in the pressure range $0-23$ GPa.
For the higher-pressure hR24 and oF16 phases, their underlying magnetic propagation vectors (if there is one) remain experimentally unknown, so we are currently excluding them in the magnetic studies.
Specifically, we utilize the rotationally invariant approach by Dudarev et al.~\cite{PhysRevB.57.1505}, which considers the following DFT+U scheme:
\begin{eqnarray}
&& E_{\text{DFT+U}} = E_{\text{LSDA}} + \\
&& \frac{(U - J)}{2} \sum_{\sigma} \left[ \left( \sum_{m_1} \hat{n}^{\sigma}_{m_1,m_1} \right) \right. \nonumber 
\left. - \left( \sum_{m_1,m_2} \hat{n}^{\sigma}_{m_1,m_2} \hat{n}^{\sigma}_{m_2,m_1} \right) \right].
\end{eqnarray}
Here, $\sigma$ is the spin index; $m_1$ and $m_2$ are orbital indices. The simplified notation $\hat{n}^\sigma_{m_1,m_2} \equiv \hat{n}_{\sigma,m_1} \hat{n}_{\sigma, m_2}$, with $\hat{n}_{\sigma,m}$ a fermion number operator of spin $\sigma$ and orbital $m$.
In this approach, the $U$ and $J$ parameters do not enter separately; only the difference $(U-J)$ is meaningful.

Converging magnetic DFT+U calculations can be challenging, especially when the energy difference between different magnetic states is small. 
We follow the recommendations of VASP developers by splitting DFT+U into four distinct, spin-polarized calculations: 
(1) Non-self-consistent initialization of magnetic moments with ENCUT = 300 eV;
(2) Non-self-consistent magnetic calculation with ENCUT =  650 eV;
(3) Self-consistent calculation with the Conjugate Gradient algorithm and a step size of 0.05;
(4) Self-consistent calculation with the Conjugate Gradient algorithm, a 0.05 step size, and inclusion of the Hubbard $U$ term.
Each consecutive calculation utilizes the wavefunction from the previous step as the starting wavefunction for the next calculation. Using this methodology, we were able to compute the energies for various magnetic configurations, including paramagentic (PM), ferromagnetic (FM), and antiferromagnetic (AFM) states.
As discussed below, our DFT+U calculations indicate an FM ground state in the hcp phase, but nontrivial AFM orderings in the $\alpha$-Sm and dhcp phases, which are consistent with the experimental findings.

\section{Results and Discussion}\label{results}

\begin{figure}[ht!]
\centering
\includegraphics[width=\linewidth]{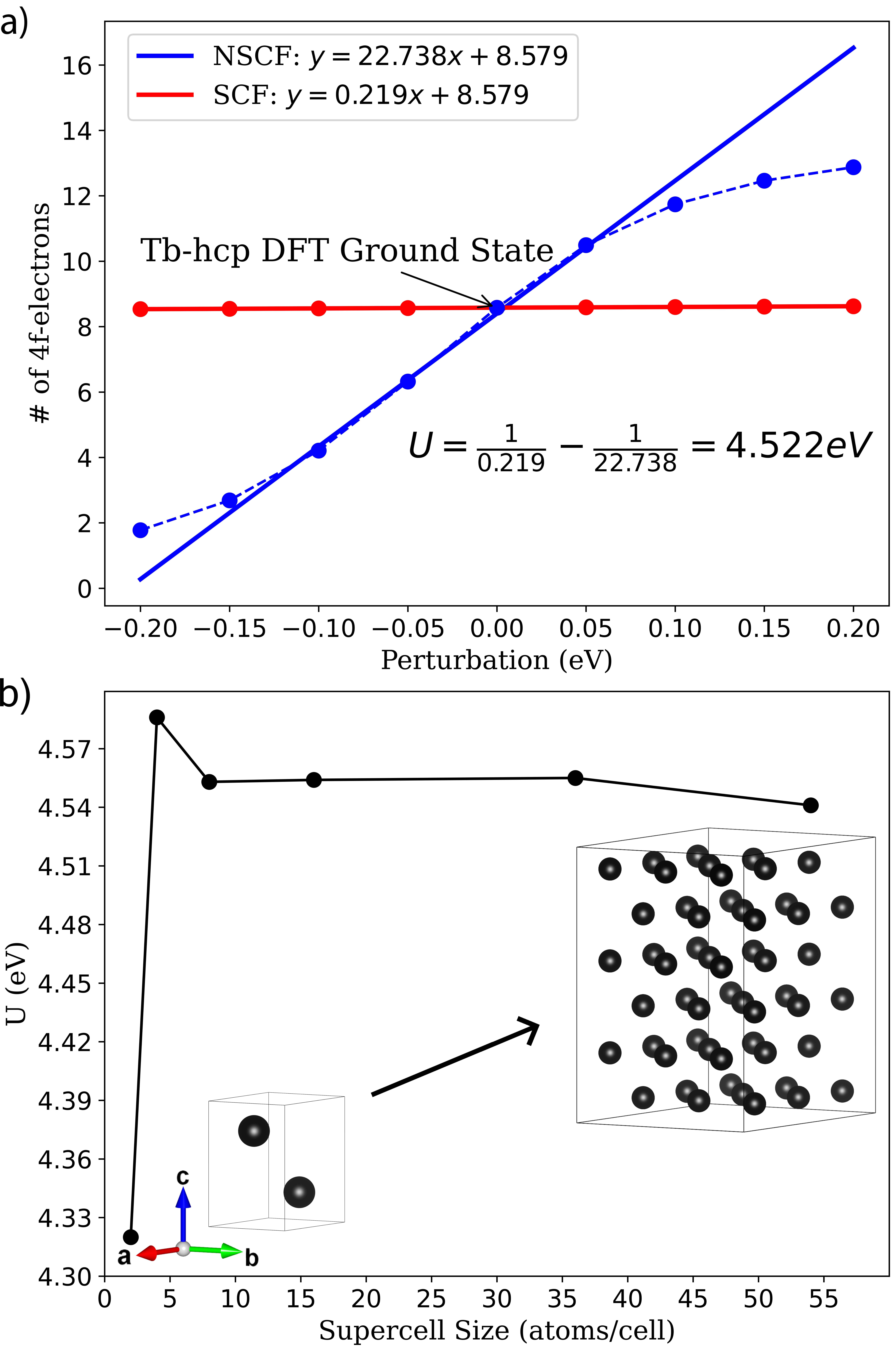}
\caption{Linear response calculations for the Hubbard $U$ parameter of Terbium (Tb). (a) $4f$-electron occupation versus applied perturbation potential, for a 16-atom supercell using the experimental crystal structure at 0.70 GPa. The inverse slopes of the fitted charge self-consistent (solid red line) and non-self-consistent (solid blue line) linear response functions determine the $U$ value.
(b) Convergence test of $U$ versus the supercell size. The results converge to $U \sim 4.55$ eV in a 54-atom Tb supercell at 0.70 GPa.}
\label{fig:LinearResponse} 
\end{figure}

Figure \ref{fig:LinearResponse}(a) depicts our linear-response calculation performed on a 16-atom Tb unit cell corresponding to the experimental hcp structure at 0.70 GPa. In this calculation, we first determine the DFT ground state. Subsequently, an external perturbation potential $V$ is applied, and the total charge occupation in the $4f$ orbitals, $n_f$, is evaluated under both self-consistent (SCF) and non-self-consistent (NSCF) conditions. Specifically, the SCF (NSCF) charge response function, $\chi \equiv \partial n_f^{SCF}/\partial V$ ($\chi_0 \equiv \partial n_f^{NSCF}/\partial V$), enables (prevents) charge density updates from the DFT ground state. The resulting Hubbard $U$ is then obtained from the difference in the inverse of the response functions~\cite{cococcioni2005linear}: $U = \chi^{-1}-\chi^{-1}_0$. When the perturbation $V$ is small, both response functions show a linear trend, enabling their direct derivation from the linear slopes in Fig. \ref{fig:LinearResponse}(a). However, with a more significant perturbation, the responses may exhibit nonlinear tendencies. Nonetheless, the linear response methodology remains applicable by employing a nonlinear fit and subsequently computing derivatives near zero potential~\cite{lambert2023use}. The result is essentially equivalent to that of a linear fit within a smaller perturbation regime. The fitting procedure leads to $U\sim 4.52$ eV in the 16-atom supercell.

The effective $U$ parameter is directly influenced by charge screening. Therefore, it is necessary to conduct a convergence test of the $U$ value as a function of the supercell size in the calculation. As shown in Fig. \ref{fig:LinearResponse}(b), the $U$ value for the Tb hcp structure at 0.70 GPa converges to $U\sim 4.55$ eV in the largest 54-atom supercell under study. For the $U$ values computed at other pressure points, we focus on supercells of 36 atoms for the $\alpha$-Sm and dhcp phase, and 48 atoms for hR24 and oF16.

\begin{figure*}[ht!]
\centering
\includegraphics[width=\textwidth]{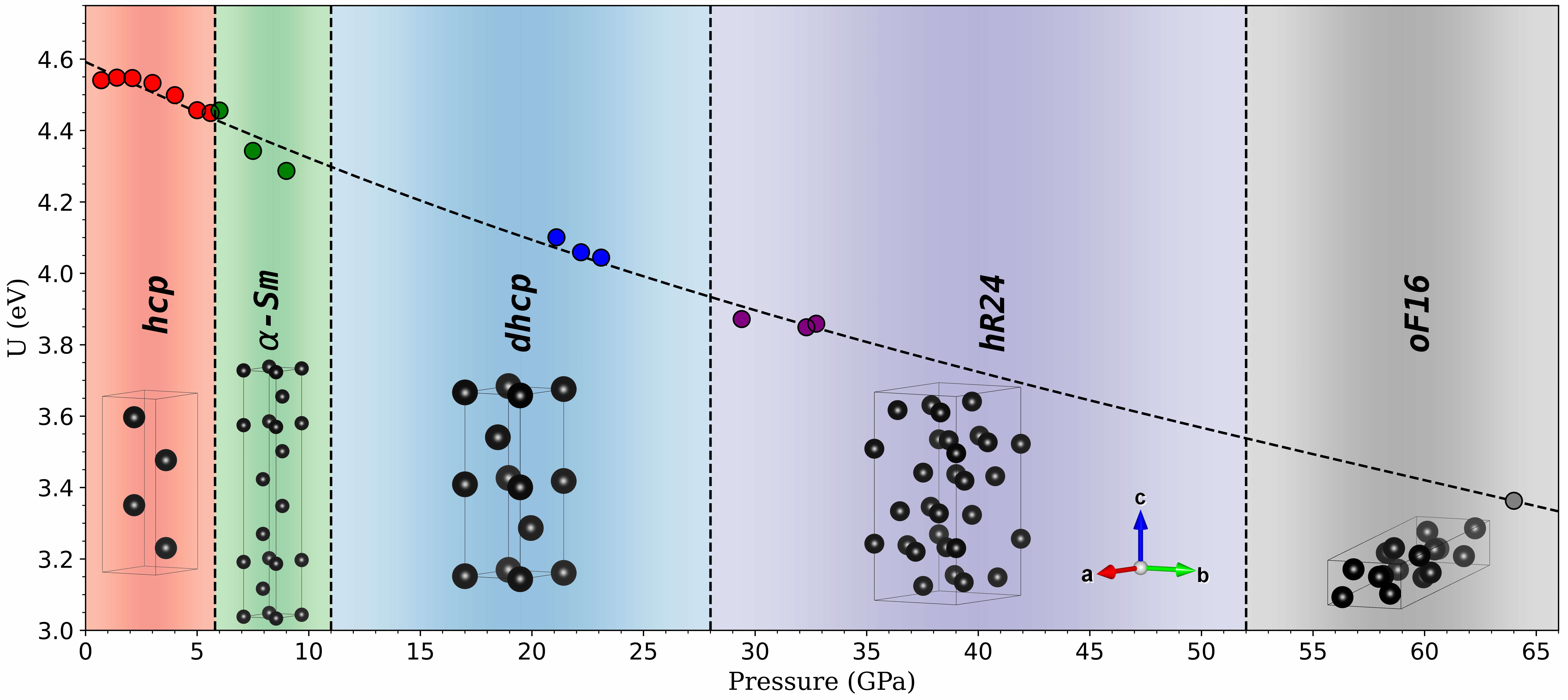}
\caption{Pressure evolution of the linear-response Hubbard $U$ parameter of Tb across different structural phase transitions. 
The vertical dashed lines indicate roughly the phase transition boundaries.
The crystal structures in the calculation are based on high-pressure experiments~\cite{kozlenko2021pressure, clay_2023, mcmahon2019structure}. The unit cells of these structures are depicted as figure insets.
The dashed curve is a simple fit for $U$ versus pressure $P$ in the range $\sim 0-65$ GPa: $U(P) = 0.0001176P^{2} - 0.02656P^{1} + 4.584$ eV, with $P$ in GPa.
}
\label{fig:UvsP}
\end{figure*}

Figure \ref{fig:UvsP} displays the $U$ values for Tb in the pressure range $P \sim 0-65$ GPa. The crystal structures in the calculation were obtained from high-pressure experiments~\cite{kozlenko2021pressure,clay_2023,mcmahon2019structure}, and these structures are depicted as figure insets representing different phases. In total, we computed the $U$ values at 17 different pressure points.
The resulting $U$ values and corresponding lattice parameters are summarized in Table \ref{table:structures}. The relationship between $U$ and $P$ can be approximately fitted by a simple function $U(P) = 0.0001176P^{2} - 0.02656P^{1} + 4.584$ eV within the pressure range under study, with $P$ in GPa. Generally, the $U$ value decreases with increasing pressure, indicating enhanced charge screening. This decrease of $U$ with $P$ is anticipated due to the reduction in bond length and the associated increase in metallicity under compression.

In the lower-pressure phases, $U$ can decline relatively faster with the applied pressure. Near the hcp to $\alpha$-Sm transition, there is a particularly sharp change in the slope of $U$ versus $P$. This change might be associated with a more dramatic structural rearrangement of atoms in the $\alpha$-Sm phase. Subsequent phases each show a progressive flattening of the slope. In the higher-pressure phases, $U$ tends to level off, indicating that the influence of pressure on the change in electron correlation effect is more minor. These $U$ versus $P$ behaviors are consistent with the expectation that it becomes increasingly more difficult to compress the crystals at higher pressure; i.e., the rate of volume reduction is smaller at higher pressure.

\begin{table}[htbp]
\centering
\begin{tabular}{@{}cccccc@{}}
\toprule
\textbf{Pressure (GPa)} & \textbf{Phase} & \textbf{a (\AA)} & \textbf{b (\AA)} & \textbf{c (\AA)} & \textbf{U (eV)} \\ \midrule
0.7                     &                & 3.5797           & 3.5797           & 5.6838           & 4.541            \\
1.4                     &                & 3.5601           & 3.5601           & 5.6512           & 4.548            \\
2.1                     &                & 3.5406           & 3.5406           & 5.6128           & 4.541            \\
3.0                     & hcp            & 3.5155           & 3.5155           & 5.5768           & 4.533            \\
4.0                     &                & 3.4876           & 3.4876           & 5.5302           & 4.499            \\
5.0                     &                & 3.4597           & 3.4597           & 5.4837           & 4.457            \\
5.6                     &                & 3.4430           & 3.4430           & 5.4357           & 4.449            \\ \hline
6.0                     &                & 3.4070           & 3.4070           & 24.9200          & 4.456            \\
7.5      & $\alpha$-Sm      & 3.3750      & 3.3750           & 24.6400          & 4.343            \\
9.0                     &                & 3.3430           & 3.3430           & 24.3600          & 4.287            \\ \hline
21.1                    &                & 3.1880           & 3.1880           & 10.2888          & 4.101            \\
22.2                    & dhcp           & 3.1835           & 3.1835           & 10.2105          & 4.059            \\
23.1                    &                & 3.1528           & 3.1528           & 10.1883          & 4.044            \\ \hline
29.4                    &                & 6.1167           & 6.1167           & 14.9613          & 3.872            \\
32.3                    & hR24           & 6.0572           & 6.0572           & 14.7356          & 3.849            \\
32.74                   &                & 6.0269           & 6.0269           & 14.8644          & 3.859            \\ \hline
64                      & oF16           & 17.9500          & 4.9330           & 2.8990           & 3.363            \\ \bottomrule
\end{tabular}
\caption{Lattice parameters of different high-pressure experimental crystal structures~\cite{kozlenko2021pressure, clay_2023, mcmahon2019structure} and the Hubbard $U$ values from first-principles linear-response calculations in this study.}
\label{table:structures}
\end{table}

In the literature, a broad spectrum of empirical $U$ values $\sim 3 - 9$ eV has been considered for Tb materials~\cite{lebegue2006multiplet,peters2014treatment,peters2016magnetism, locht2016standard,mcmahon2019structure}. Here, our first-principles linear-response calculations have narrowed down the range to $U\sim 4.5 - 3.5$ eV in the pressure range $\sim 0- 65$ GPa. Below, we demonstrate that these $U$ values in self-consistent DFT+U calculations lead to magnetic ground states in the hcp, $\alpha$-Sm, and dhcp phases, with magnetic ordering vectors compatible with the experimental findings. 

\begin{figure}[ht!]
\centering
\includegraphics[width=\linewidth]{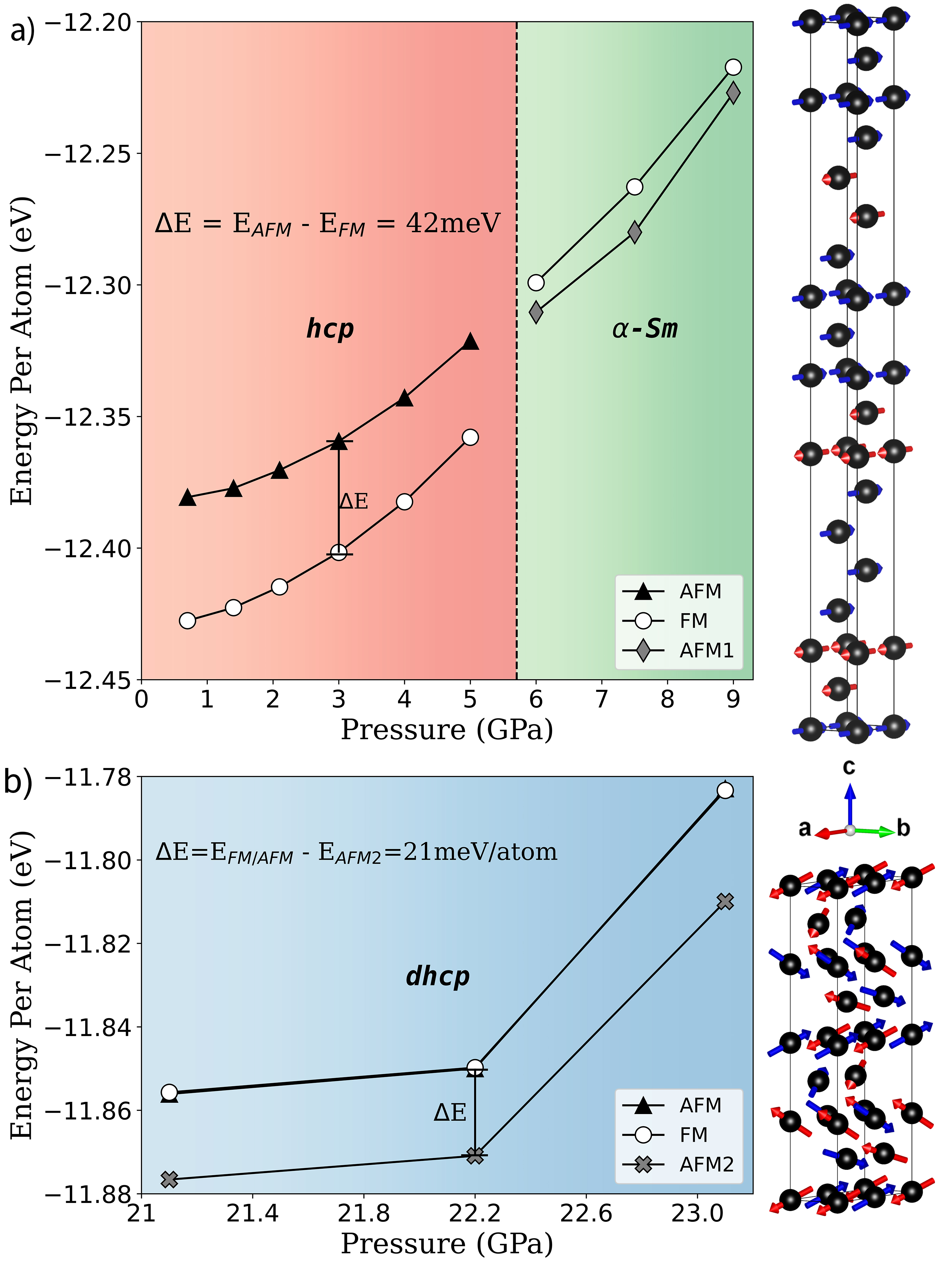}
\caption{Energy per atom computed by DFT+U for different magnetic phases of Tb in the pressure range $0-23$ GPa. The calculations confirm that the ferromagnetic (FM) state is suppressed at high pressures. (a) The left (red) panel is for the hcp phase, where the  FM state has the lowest energy. The right (green) panel is for the $\alpha$-Sm phase, where an antiferromagnetic (AFM) configuration denoted as AFM1 with a magnetic propagation vector $k_{\textrm{AFM1}}=(0, 0, \frac{1}{2})$ has the lowest energy. This AFM1 state proposed by Kozlenko et al.~\cite{kozlenko2021pressure} is shown in the right inset.
(b) The lowest-energy state in the dhcp phase has a magnetic configuration denoted as AFM2 with a propagation vector $k_{\textrm{AFM2}}=(\frac{1}{2}, 0, \frac{1}{2})$ shown in the right inset. Our AFM2 configuration exhibits a complex, non-collinear spin arrangement.}
\label{fig:EnergyvsP} 
\end{figure}

Figure \ref{fig:EnergyvsP} shows the energy per atom computed by DFT+U for Tb metal in different magnetic states, using the resulting $U$ values from our linear-response calculations. The calculations, performed in the pressure range of $0-23$ GPa, cover the hcp, $\alpha$-Sm, and dhcp structures. The input crystal structures in the calculation were obtained from corresponding high-pressure experiments~\cite{kozlenko2021pressure,clay_2023}.
The hR24 and oF16 phases are omitted here, as the underlying magnetic propagation vectors (if there is one) remain experimentally unknown.

In Fig. \ref{fig:EnergyvsP}, the (collinear, inter-layer) AFM state has a unit cell identical to that in the PM or FM phase. In contrast, the AFM1 and AFM2 states have enlarged unit cells, with magnetic ordering vectors $k_{\textrm{AFM1}}= (0,0,\frac{1}{2})$ and $k_{\textrm{AFM2}}= (\frac{1}{2},0,\frac{1}{2})$, respectively. Compared to a non-magnetic state, the unit cell in AFM1 is doubled along the crystallographic $c$-axis, while in AFM2, it is enlarged by a factor of two along both the $a$- and $c$-axes.

As shown in Fig. \ref{fig:EnergyvsP}(a), the lowest-energy state in the low-pressure hcp phase is FM. However, across the transition to the $\alpha$-Sm phase, an AFM1 configuration proposed by Kozlenko et al.~\cite{kozlenko2021pressure} has a lower energy. This AFM1 state with an in-plane spin alignment is shown schematically in Fig. \ref{fig:EnergyvsP}(a) inset.
In the higher-pressure dhcp phase, an AFM2 configuration depicted in Fig. \ref{fig:EnergyvsP}(b) inset becomes the lowest-energy state. Neutron diffraction studies have previously observed a pressure-induced structural phase transition in Tb metal from hcp to $\alpha$-Sm near 4 GPa, accompanied by a magnetic transition from FM to an AFM order~\cite{kozlenko2021pressure}. Our DFT+U results agree with the experiments, indicating a lower energy FM state in the low-pressure hcp phase, and preferred AFM1 and AFM2 states in the higher-pressure $\alpha$-Sm and dhcp phases, respectively~\cite{kozlenko2021pressure, clay_2023}.

We note that the AFM2 configuration for the dhcp phase is geometrically complex, involving a non-collinear spin arrangement. In our iterative self-consistent calculations, although the initial spin moments were restricted to only in-plane directions, the calculation ended up converging to an AFM2 state with tilted spins arranged in a non-collinear manner [Fig. \ref{fig:EnergyvsP}(b) inset].
Specifically, for atoms aligned vertically along the edges of the unit cell, their spin directions follow a repeating angle of rotation. Starting from the 2a \((0,0,0)\) Wyckoff position and moving upwards along the $c$-axis, the angles of spin rotation between consecutive atoms are roughly \(63^\circ\), \(117^\circ\), \(63^\circ\), \(117^\circ\). These spins located at the 2a Wyckoff positions are co-planar. Moving along the \(a\)-axis, there is a \(180^\circ\) rotation between each consecutive 2a position. Similarly, atoms located on the 2c Wyckoff positions at \((\frac{1}{3},\frac{2}{3},\frac{1}{4})\) and \((\frac{2}{3},\frac{1}{3},\frac{3}{4})\) form another set of coplanar spins. Beginning at 2c \((\frac{1}{3},\frac{2}{3},\frac{1}{4})\) and then alternating to 2c \((\frac{2}{3},\frac{1}{3},\frac{3}{4})\) as one travels up along the $c$-axis, the angles of rotation repeat as \(24^\circ\), \(-24^\circ\), \(24^\circ\). Moving along the \(a\)-axis, there is again an alternating \(180^\circ\) rotation between adjacent 2c Wyckoff sites.
It would be an important future study to see if such a non-collinear magnetism can be detected experimentally in the high-pressure dhcp phase.

\begin{figure}[ht!]
\centering
\includegraphics[scale=0.46, width=\linewidth]{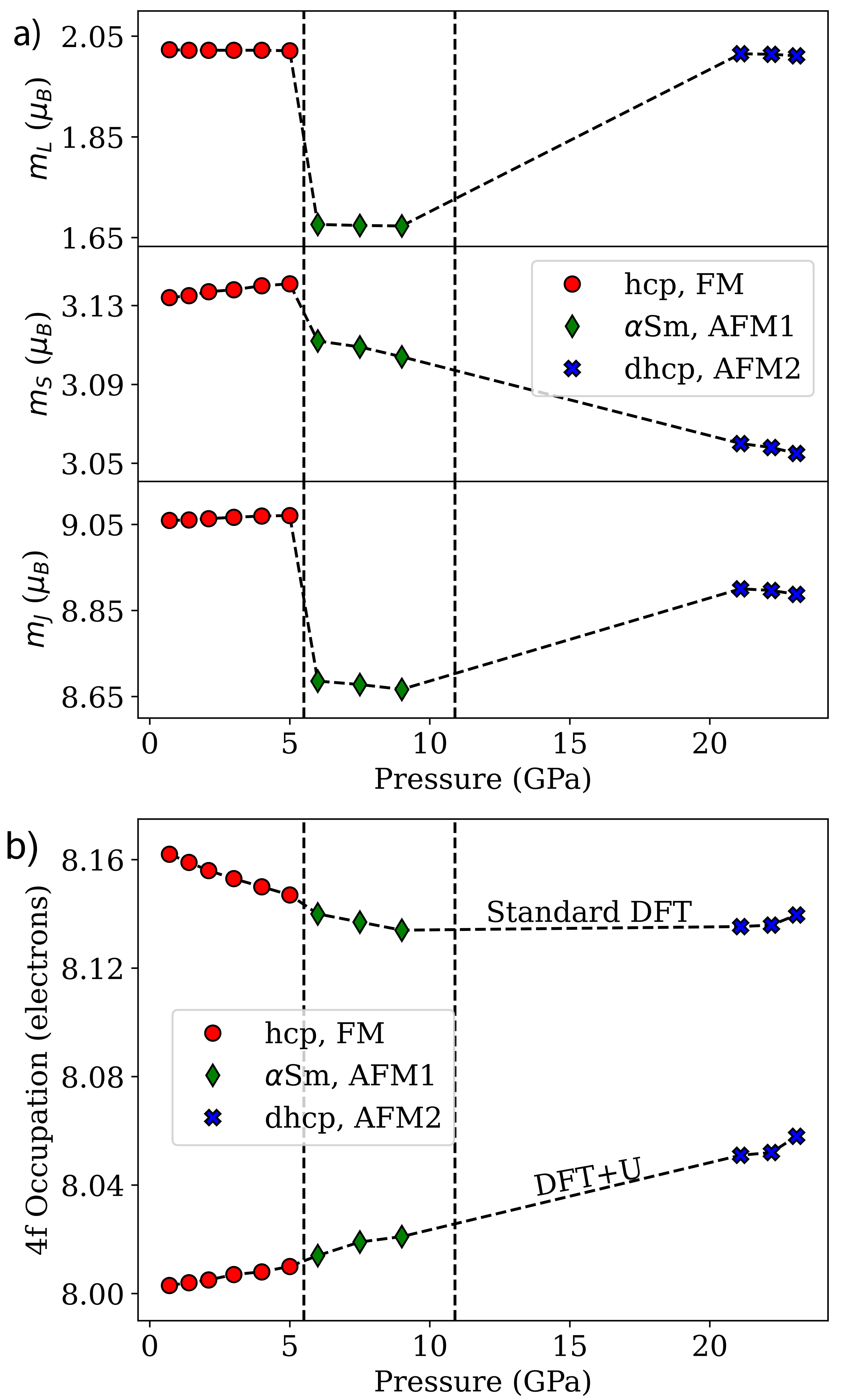}
\caption{Pressure evolution of the magnetic moment size and charge occupation for the lowest-energy states in our DFT+U calculations respectively for the hcp, $\alpha$-Sm, and dhcp phases.
(a) The moments are decomposed into the orbital ($m_L$), spin ($m_S$), and total angular momentum ($m_J$) contributions, measured in Bohr magneton ($\mu_B$).
(b) Number of $4f$ electrons from both standard DFT and DFT+U calculations. The DFT+U result exhibits a lower $4f$-orbital occupation due to an additional on-site Hubbard repulsion term.}
\label{fig:MomentOccupation} 
\end{figure}

Finally, Fig. \ref{fig:MomentOccupation} shows the averaged Tb moment size and the $4f$-orbital occupation of the lowest-energy states for DFT+U calculations in the pressure range $0-23$ GPa. In Fig. \ref{fig:MomentOccupation}(a), the moments are decomposed into the orbital ($m_L$), spin ($m_S$), and total angular momentum ($m_J$) contributions, respectively. The total moment size is $\sim 9$ $\mu_B$, which is consistent with the Hund's rule ground state of a Tb $4f^8 (5d6s)^3$ electronic configuration.
As shown in Fig. \ref{fig:MomentOccupation}(b), the DFT+U $4f$ occupation is indeed close to 8. For a comparison, the standard DFT result (without $U$) is also shown in the figure. 
DFT+U generally predicts fewer electrons in the $4f$ orbitals compared to standard DFT for all studied pressures.
As anticipated, in the presence of a Hubbard repulsion, double occupancy in $4f$ orbitals is discouraged by electron correlation effects, favoring the hybridization of $4f$ electrons with $5d6s$ conduction bands.
The charge occupation information will be important for other quantum many-body techniques, such as dynamical mean-field theory or exact diagonalization working in a truncated Hilbert space.

\section{Conclusion}\label{conclusion}
In summary, we utilized first-principles linear-response calculations to determine the $4f$-orbital Hubbard parameter $U$ for Tb metal under pressure. We found that $U$ ranges from 4.5 to 3.5 eV in the pressure range $0-65$ GPa. We further computed the energies for different high-pressure magnetic phases, using self-consistent DFT+U approaches with the linear-response $U$ values and experimental crystal structures at corresponding pressure points.
We found that the lowest-energy state under study is ferromagnetic (FM) in the low-pressure hcp phase. At higher pressures, antiferromagnetic (AFM) states can dominate, with magnetic propagation vectors $k_{\text{AFM1}}=(0,0,\frac{1}{2})$ and $k_{\text{AFM2}}=(\frac{1}{2},0,\frac{1}{2})$ for the $\alpha$-Sm and dhcp phases, respectively. 
These results match the experimental findings, helping validate the linear-response predictions for the pressure-dependent $U$ parameter and affirming the predictive power of DFT+U for addressing the complex interplay of structural and magnetic transitions in correlated systems.
Our $U$ values also provide valuable inputs for other quantum many-body techniques for future studies of Tb under high pressure.

We conclude by discussing potential limitations of our work and future research directions. First, DFT is essentially a zero-temperature technique, so it cannot capture, for example, the additional finite-temperature helical-AFM to FM transition in the hcp phase with reducing temperature. 
To address this, other techniques such as dynamical mean-field theory (DMFT) would be needed. A fully self-consistent DFT+DMFT with iterative feedback between the two methods remains lacking in the literature for the studies of Tb under pressure. Second, the AFM configurations under study are naturally not exhaustive, and we cannot exclude the possibility of another lower-energy AFM state with the same magnetic propagation vector. More studies on different AFM configurations compatible with the crystal symmetry and magnetic ordering would be useful. Finally, DFT+U can be viewed as a static mean-field theory for electron interaction effect. Other advanced quantum many-body techniques capable of treating frequency and/or momentum-dependent corrections of the self energy can further help validate our study.
In particular, it would be important to check the non-collinear AFM2 state discovered in our DFT+U calculation in the dhcp phase, both theoretically and experimentally. Benchmarking our $U$ values with other techniques, such as constrained random-phase approximation, and applying these techniques to other rare-earth systems are also interesting future research areas.

\section*{acknowledgments}
This work is supported by the U.S. Department of Energy (DOE) Basic Energy Sciences (BES) Program under Award No. DE-SC0023268. L.B. also acknowledges support from the NASA Alabama Space Grant Consortium Scholarship. 
This research used resources of the National Energy Research
Scientific Computing Center (NERSC), a U.S. DOE Office of Science User Facility supported under Contract No. DE-AC02-05CH11231 using NERSC Award BES-ERCAP0028499.


\bibliography{apssamp}

\end{document}